\documentclass[aps,prl,twocolumn,showpacs]{revtex4}

\usepackage{graphicx} \usepackage{amsmath}
\usepackage[colorlinks]{hyperref}

\newcommand{\EqLabel}[1]{\label{#1}} \newcommand{\mb}[1]{\mathbf{#1}}

\begin{document}
 
\title{Few-particle Green's functions for strongly correlated
  systems on  infinite lattices}

\author{Mona Berciu} 

\affiliation{Department of Physics \&  Astronomy, University of
  British Columbia, Vancouver, BC, Canada, 
  V6T 1Z1}

\date{\today}
 
\begin{abstract} 
We show how few-particle Green's functions can be calculated
efficiently for models with nearest-neighbor hopping, for infinite
lattices in any dimension. As an example, for one dimensional spinless
fermions with both nearest-neighbor and second nearest-neighbor
interactions, we investigate the ground states for up to 5
fermions. This allows us not only to find the stability region of
various bound complexes, but also to infer the phase diagram at small
but finite concentrations. 
\end{abstract}


\maketitle


Recently, there has been considerable interest in few-particle
solutions of interacting Hamiltonians. For example, in
Ref. \cite{Drummond} it was shown that knowledge of the two- and
three-body solutions allows for quantitatively accurate predictions of
finite-temperature thermodynamic quantities for many-body systems. As
another example, in the context of atomic and molecular physics, the
predicted universal three-body Efimov structures \cite{Efimov} have
now been seen experimentally \cite{Hulet}, giving new impetus to their
study and work on various generalizations \cite{Fei}.

While the above work is for free space where particles 
 have parabolic dispersions, there is equally strong
interest in the lattice version of such few-body problems. For
example, while stable excitons -- bound pairs comprised of an electron
and a hole -- appear in many materials, it is less clear when  a
 so-called charged exciton or trion, consisting of two holes and one
 electron or 
viceversa, is stable. That
this can happen has been recently demonstrated in GaAs quantum wells
\cite{GaAs} and in carbon nanotubes \cite{cano}. (Note that trion
theory is still mostly based on continuous models and variational
solutions, {\em e.g.} see Ref. \cite{trion-the}). Studying bigger
bound complexes, for example bi-exciton pairs, is the next logical
step.

Few-particle bound states are relevant not only for the materials
where they appear, but also in the interpretation of certain
spectroscopic data. For instance, the role played by bound
two-particle states, leading to atomic-like multiplet structures in
the Auger spectra of narrow-band insulating oxides, is well
established \cite{george}. At low dopings, more complicated complexes
may form and leave their fingerprints in various spectroscopic
features. It is therefore useful to be able to study relatively easily
few-particle solutions on an infinite lattice.

In this Letter we show that few-particle Green's functions can be
calculated efficiently for strongly correlated lattice Hamiltonians in
the thermodynamic limit, at least so long as the hopping involves only
nearest neighbor sites. For simplicity and to illustrate the
technique and its usefulness, we focus here on a one-dimensional (1D)
model of spinless fermions with nearest-neighbor (nn) and
next-nearest-neighbor (nnn) interactions. However, the method
generalizes straightforwardly to higher dimensions, longer (but
finite) range interactions, mixtures of fermions (including spinful
fermions) and/or bosons, etc. Such problems are of direct interest
either in solid state physics, or for cold atoms in optical lattices.

For two-fermion Green's functions ($N_f=2$), our method is equivalent
to that of Ref. \cite{george}, but is recast in a simpler form which
allows, in 1D, for an analytical solution for any finite-range
interaction. More importantly, it has a simple generalization for
$N_f>2$. We study cases with up to $N_f=5$ and show that these suffice
not only to sort out the stability of few-particle bound states, but also
to infer the low density phase diagram.

Consider, then, spinless fermions on a 1D chain with $N\rightarrow
\infty$ sites, described 
by the Hamiltonian:
\begin{equation}
\EqLabel{1}
\nonumber
 {\cal H}=-t\sum_{i}^{} (c_i^\dag c_{i+i} + h.c.) + U_1
\sum_{i}^{} n_i n_{i+1} + U_2 \sum_{i}^{} n_i n_{i+2}
\end{equation}
where $c_i$ removes a spinless fermion from site $i$ located at $R_i = ia$,
and $n_i=c_i^\dagger c_i$. Note that this 1D Hamiltonian is not
integrable in the sense of having a Bethe ansatz solution. Because
our solution is not linked in any way to such integrability, it can be
generalized to higher dimensions, as already mentioned. To illustrate
the main idea behind our solution, we discuss in some detail the solution
for $N_f=2$ fermions, after which we generalize to $N_f>2$.  Other
possible generalizations, mentioned above, 
are discussed in the supplementary material \cite{supp}.

Because the Hamiltonian is invariant to translations, the total
momentum of the pair is a good quantum number. As a result, we work with the
$N_f=2$ states:
$$ 
|k, n\rangle = {1\over \sqrt{N}} \sum_{i}^{} e^{ik
  \left(R_i+{na\over 2}\right)} c_i^\dag c^\dag_{i+n}|0\rangle
$$ which describe  fermions at a relative distance $n\ge 1$.

We define the two-particle Green's functions:
\begin{equation}
\nonumber \EqLabel{2} G(m, n; k, \omega) = \langle k, m |
\hat{G}(\omega) | k, n\rangle 
\end{equation}
where $\hat{G}(\omega) = [\omega +i \eta - {\cal H}]^{-1}$ with
    $\eta\rightarrow 0_+$ and we set $\hbar=1$. From the Lehmann
    representation:
$$ G(m, n; k, \omega) = \sum_{\alpha}^{} \frac{\langle k, m |
  k,\alpha\rangle\langle k,\alpha| k, n\rangle}{\omega -
  E_{2,\alpha}(k) + i \eta},
$$ where $\{ |k,\alpha\rangle\}$ are the two-particle eigenstates with
total momentum $k$, ${\cal H}|k,\alpha\rangle =
E_{2,\alpha}(k)|k,\alpha\rangle$. Thus, this propagator allows us to
find the $N_f=2$ spectrum and also to get information about its
eigenfunctions. Its Fourier transform $G(m,n;k,t)\propto \langle k, m
| \exp(-i {\cal H} t)| k, n\rangle$ is the amplitude of probability
that if initially the two particles (with total momentum $k$) are at a
relative distance $na$, they will be at a relative distance $ma$ after
time $t$.

Matrix elements of the
  identity $1={\hat G}(\omega) \left( \omega+ i \eta- {\cal H}\right)$ lead to
$ \delta_{n,m} = (\omega +i \eta) G(m, n; k, \omega) - \langle k, m |
\hat{G}(\omega){\cal H} | k, n\rangle$. Since
${\cal H} | k, n\rangle=  U(n) |k, n\rangle- f(k)\left[|k, n-1\rangle +
|k,n+1\rangle\right]$, where $U(n) = U_1 \delta_{n,1} + U_2 
\delta_{n,2}$ and $f(k) = 2t \cos {ka\over 2}$, we get a  simple
  recurrence relation:
\begin{multline}
 \delta_{n,m} =  [\omega + i \eta - U(n)] 
G(m,n; k, \omega) \\+ f(k) \left[G(m,n-1;k,\omega) +
G(m,n+1;k,\omega)\right]
\end{multline}
This is  trivial to solve for an infinite chain if one realizes that
 for any $m$ of interest, $G(m,n; k, \omega)
 \rightarrow 0$ as  $n\rightarrow\infty$. This is obvious if $\omega$ is outside the
  free two-particle continuum where eigenstates, if any, are bound
  and therefore wavefunctions decay exponentially with $n$. It is also
  true inside the free two-particle continuum. Even though here the
  wavefunctions are plane-waves,  $\eta$
  defines an effective lifetime $\tau \sim 1/\eta$. As such,
  $G(m,n; k, t)\rightarrow 0$ if $na$ is large compared to
  the typical distance that  particles 
  travel within $\tau$. Thus, the recurrence relation can be
  solved starting from $G(m,M_c+1; k, \omega)=0$ for a sufficiently
  large cutoff $M_c$. Of course, the $N_f=2$ case can be solved analytically
  exactly (see below). However, the idea can be used for $N_f>2$
  cases, where a numerical solution is needed. Noting that the
  few-particle Green's functions become arbitrarily small as a
  ``relative distance'' $M$ (to be 
  defined below) increases, the recurrence relations can be
  solved propagating the solution from a cutoff $M_c$ towards
  small $M$. $M_c$ is then increased until 
  convergence is reached. The effects of $\eta$ and
 $M_c$ on the numerical solution are discussed in the supplementary material.

First, though, we  complete the  $N_f=2$ discussion, which has an
analytical solution (for 
details see \cite{supp}; we also show there how to deal with a
finite-size system in this case).  At the Brillouin zone (BZ) edge, since
$f(k=\pi/a)=0$  we find:
$$ 
G(1,n;{\pi\over a},\omega)={\delta_{n,1}\over \omega+i\eta -U_1},
$$ 
as expected since $|{\pi\over a}, 1\rangle$ is an eigenstate of ${\cal
  H}$ with energy $U_1$.  For any
$ka\ne \pi$, we find
$$ 
G(1,1;k,\omega) = \left[\omega + i \eta - U_1 -
  \frac{[f(k)]^2}{\omega + i \eta -U_2 + z(k,\omega)f(k)}\right]^{-1}
$$ and for any $n\ge 2$, 
$$
G(1,n;k,\omega) =- \frac{[z(k,\omega)]^{n-1}f(k)
  G(1,1;k,\omega)}{\omega + i \eta -U_2 + z(k,\omega)f(k)}.  $$
Values for $m>1$ can be obtained similarly.
Here, $z(k,\omega)$ is the root of the characteristic equation
of  this recurrence relation, $
(\omega + i \eta) + f(k)\left( z+{1\over z}\right)=0$, for which
$|z(k,\omega)|<1$ \cite{supp}. This shows that indeed, $G(1,n;,k,\omega)
\rightarrow 0$ as $n\rightarrow \infty$. It is also easy to check
that inside the free
two-particle continuum, $|\omega| < 2 f(k)$, we have $1-|z(k,\omega)|\sim
\eta$, so here $G$ decays exponentially only because $\eta >0$.

\begin{figure}[t]
\includegraphics[width=0.95\columnwidth]{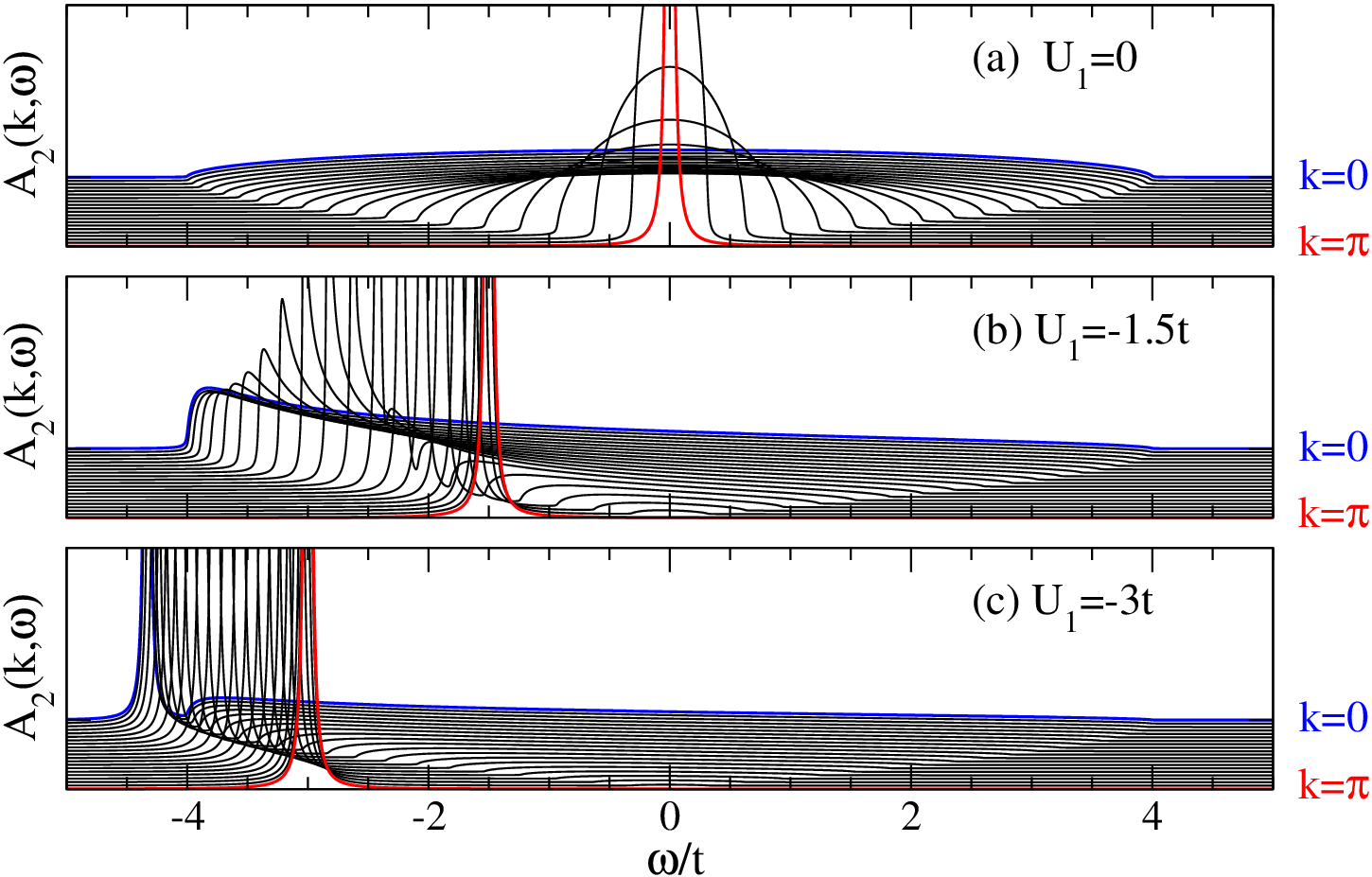}
\caption{(color online) $A_2(k,\omega)$ for
  $U_2=0$ and 
  (a) $U_1=0$; (b) $U_1=-1.5t$ and (c) $U_1=-3t$. As the nn attraction
  is turned on, a bound state splits from the free two-particle continuum
  shown in (a). It exists at all momenta if $U_1 <-2t$, but only for large
  momenta if $U_1> -2t$. Here $\eta=0.01$.
\label{fig1}}
\end{figure}

To study the two-particle spectrum, we plot the two-particle spectral weight:
$A_2(k,\omega) = -{1\over \pi} \mbox{Im} G(1,1;k,\omega)$ in
Fig. (\ref{fig1}) for $U_2=0$, and three values of $U_1$.  By
definition, $A_2(k,\omega)$ is
finite at energies in the two-particle spectrum, and its value is
related to the probability to find the fermions as nn in that eigenstate.
If $U_1=0$,
$A_2(k,\omega)$ is finite in the free two-particle continuum, ranging
from $-4t$ to $ 4t$ if $k=0$, while at $k=\pi/a$ only $\omega=0$ is an
eigenstate, hence the $\delta$-function (Lorentzian) seen here.  As an
attractive $U_1$ is turned on, the $k=\pi/a$ peak tracks $U_1$, and a
bound state is pulled below the continuum at nearby $k$ values. For
$U_1 > -2t$, this bound state exists only near the BZ edge, while near
the $\Gamma$ point the weak attraction shifts spectral weight to the
bottom of the two-particle continuum but is not enough to push a
discrete state
below it. For $U_1 < -2t$, the bound state
becomes the low-energy state at all $k$. This shows that, for certain
ranges of parameters, bound pairs are only stable in some regions of
the BZ, which moreover are not necessarily near $k=0$. It would be
interesting to investigate their effects on various response
functions.

\begin{figure}[t]
\includegraphics[width=0.8\columnwidth]{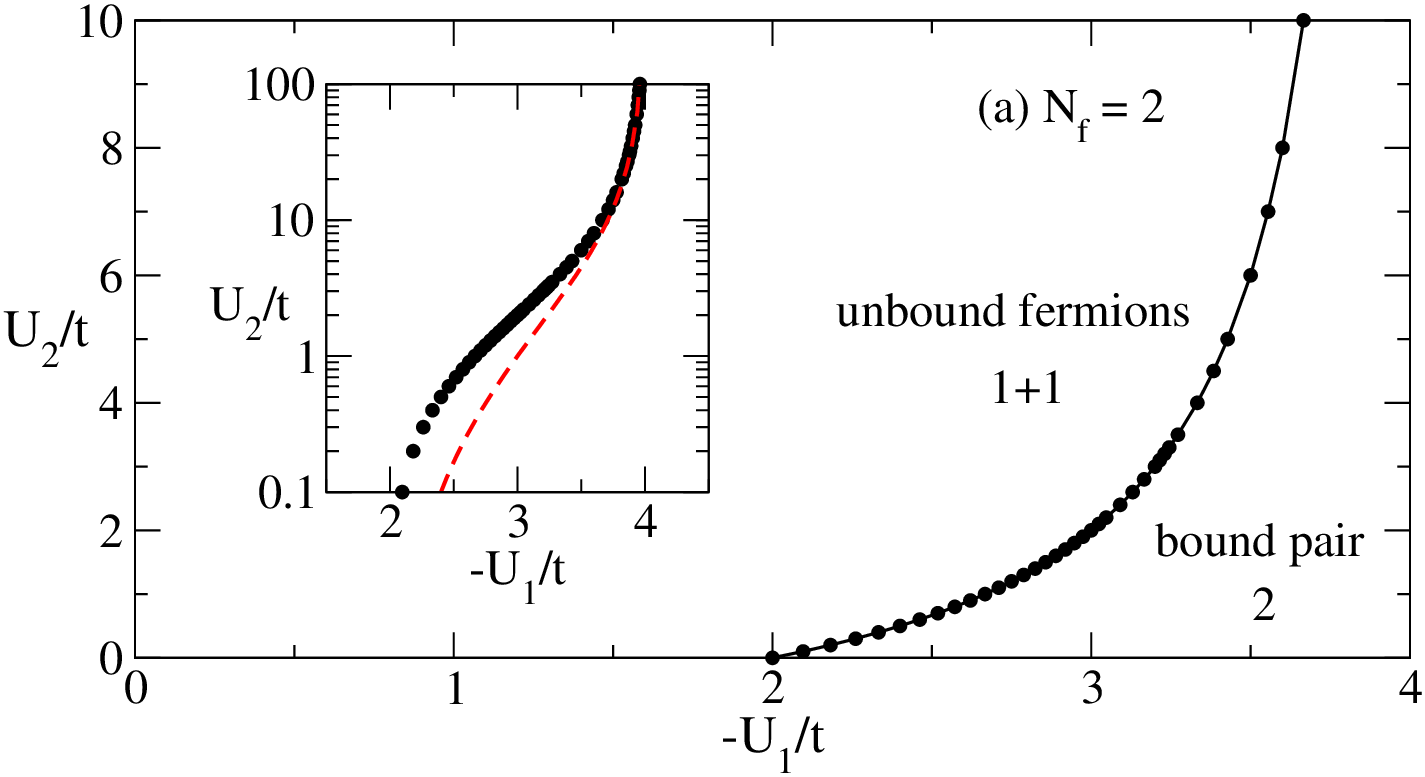}
\includegraphics[width=0.8\columnwidth]{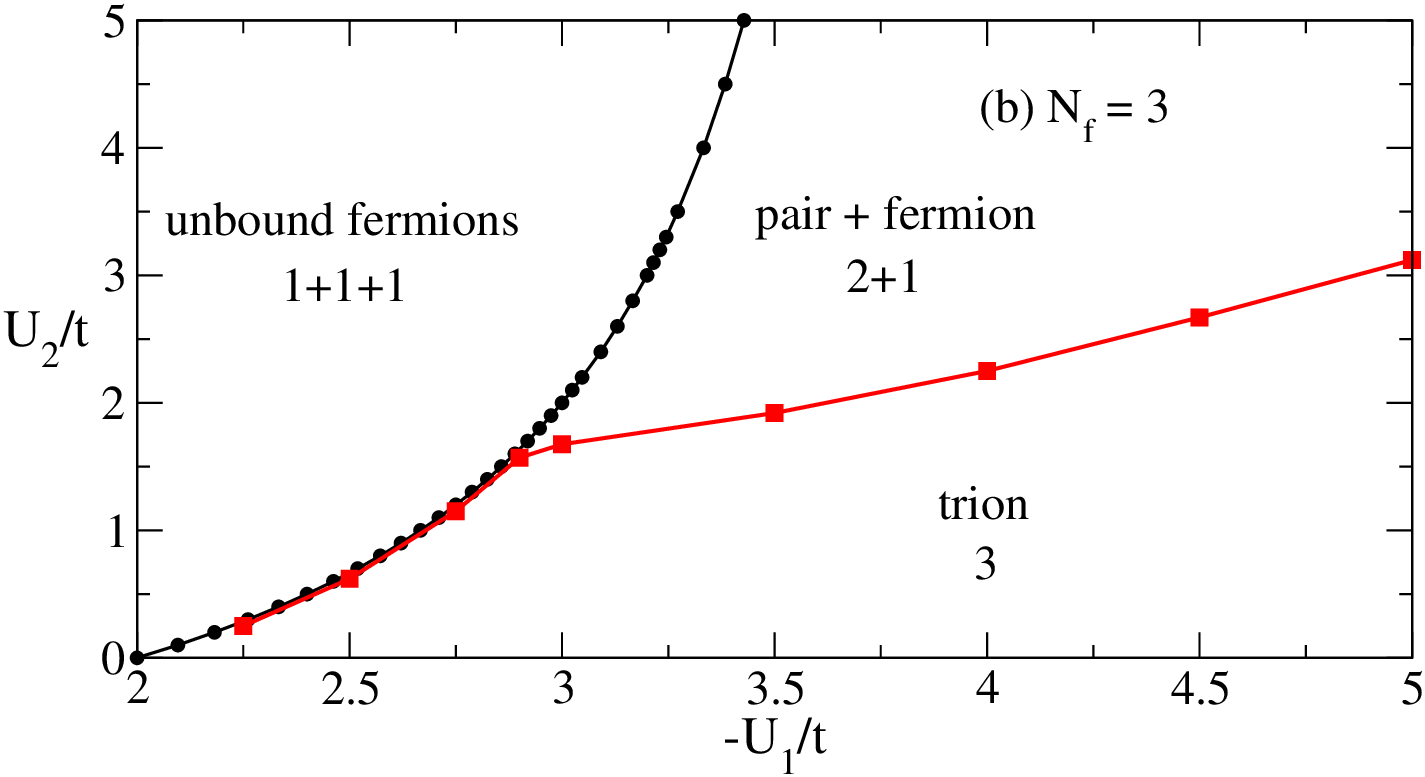}
\caption{(color online) Stability diagram for (a) $N_f=2$, and (b)
  $N_f=3$ fermion systems, indicating the nature of the GS. The inset
  in (a) shows that bound pairs are always stable if $U_1<-4t$. The
  dashed line is a perturbational prediction.
\label{fig2}}
\end{figure}

However, hereafter we focus on the $k=0$ ground-state
(GS). Fig. \ref{fig2}a shows  whether in the
GS the pair is bound or not, for $U_1<0$ and $U_2>0$. (Note that such
interactions, attractive at 
short-range and repulsive at longer-range, appear in systems with
highly polarizable ions \cite{pnic}).  For $U_1
<-4t$ a bound pair is always stable; even if it had infinite mass, a
nn  pair of energy $U_1$ is below the minimum energy of two
free fermions, of $-4t$. Of course, the kinetic energy of the
pair further enhances its stability region. The full line in the inset
shows a perturbational estimate for $t\ll |U_1-U_2|$ \cite{supp}.

This $N_f=2$ stability diagram, however, has no predictive power for
what happens if more fermions are in the system. For example, if
$N_f=3$, we expect regions where the GS consists of 3  fermions,
of a bound pair plus a  fermion, or of a bound ``trion''. To
identify these regions we study $N_f=3$ Green's functions, by direct
generalization of the $N_f=2$ approach. Briefly, for any $n_1\ge 1,
n_2\ge 1$, we define three-particle states:
$$ |k, n_1, n_2\rangle = {1\over \sqrt{N}} \sum_{i}^{} e^{ikR_i}
c^\dag_{i-n_1} c^\dag_i c^\dag_{i+n_2}|0\rangle
$$ and three-particle Green's functions:
$$ G(m_1, m_2; n_1, n_2; k, \omega) = \langle k, m_1, m_2|
\hat{G}(\omega)| k, n_1, n_2\rangle.
$$ Recurrence relations for these propagators are generated just as
for the $N_f=2$ case. If we define a ``relative distance''
$M=n_1+n_2$, hopping of the outside fermions will link Green's
functions with a given $M$ to those with $M\pm 1$.  If the central
fermion hops, one of the $n_1,n_2$ values increases by one and the other
decreases by one, therefore $M$ remains the same. Thus, the equation
of motion links Green's functions with consecutive $M-1,M,M+1$ values,
leading to recurrence relations that can be solved in terms of
continued fractions of matrices, if we use the insight that
propagators vanish as $M\rightarrow \infty$. Generalization to larger
$N_f$ values is
now straightforward \cite{supp}.

In higher dimension, we need to combine the ``relative distance'' with the
``Manhattan distance'' 
\cite{mash}. For example, in 2D for $N_f=3$, we associate the
plane-wave with the coordinates $i_x$ and $i_y$ of the ``central''
particle for that axis. The other particles' coordinates are
$i_x-n_{1,x}$, $i_x+n_{2,x}$, respectively $i_y-n_{1,y}$,
$i_y+n_{2,y}$, where $n_{i,\alpha}\ge 0$, $i=1,2, \alpha=x,y$.  If we
choose $M=\sum_{i,\alpha}^{}n_{i,\alpha}$ then nn hopping links
together only Green's functions with $M-1, M, M+1$. Thus, $m$
particles in 2D is computationally similar to $2m-1$ particles in 1D.
In both cases, $2(m-1)$ positive integers specify the relative positions, and
$M$ is their sum. The key observation is that the equations of
motion still group into recurrence relations linking only
quantities with $M-1,M,M+1$, allowing for an efficient solution  (for more details, see \cite{supp}).

To study the spectrum of the $N_f=3$, 1D system, we plot 
$A_3(k,\omega)= -{1\over \pi} \mbox{Im} G(1,1;1,1;k,\omega)$. This
must have finite spectral weight for $\omega \ge E_{2,GS}-2t$,
corresponding to a continuum of states describing a fermion 
far away from a pair. (If $E_{2,GS}=-4t$, this continuum starts at $-6t$
and describes 3 free fermions). If the continuum is the lowest
spectral feature, then the GS is either a pair+fermion or three
fermions, mirroring the $N_f=2$ situation. However, if a discrete state
appears below this continuum, 
then the GS is a stable bound trion \cite{supp}. The stability  diagram is
plotted in Fig. \ref{fig2}b and shows a region where trions are stable,
at large attractive $U_1$ and weak repulsive $U_2$. This is expected since
 binding a 3rd fermion to a stable pair  lowers its
energy by roughly $U_1+U_2$, while a free fermion can lower the total
energy by at most $-2t$.

\begin{figure}[t]
\includegraphics[width=0.8\columnwidth]{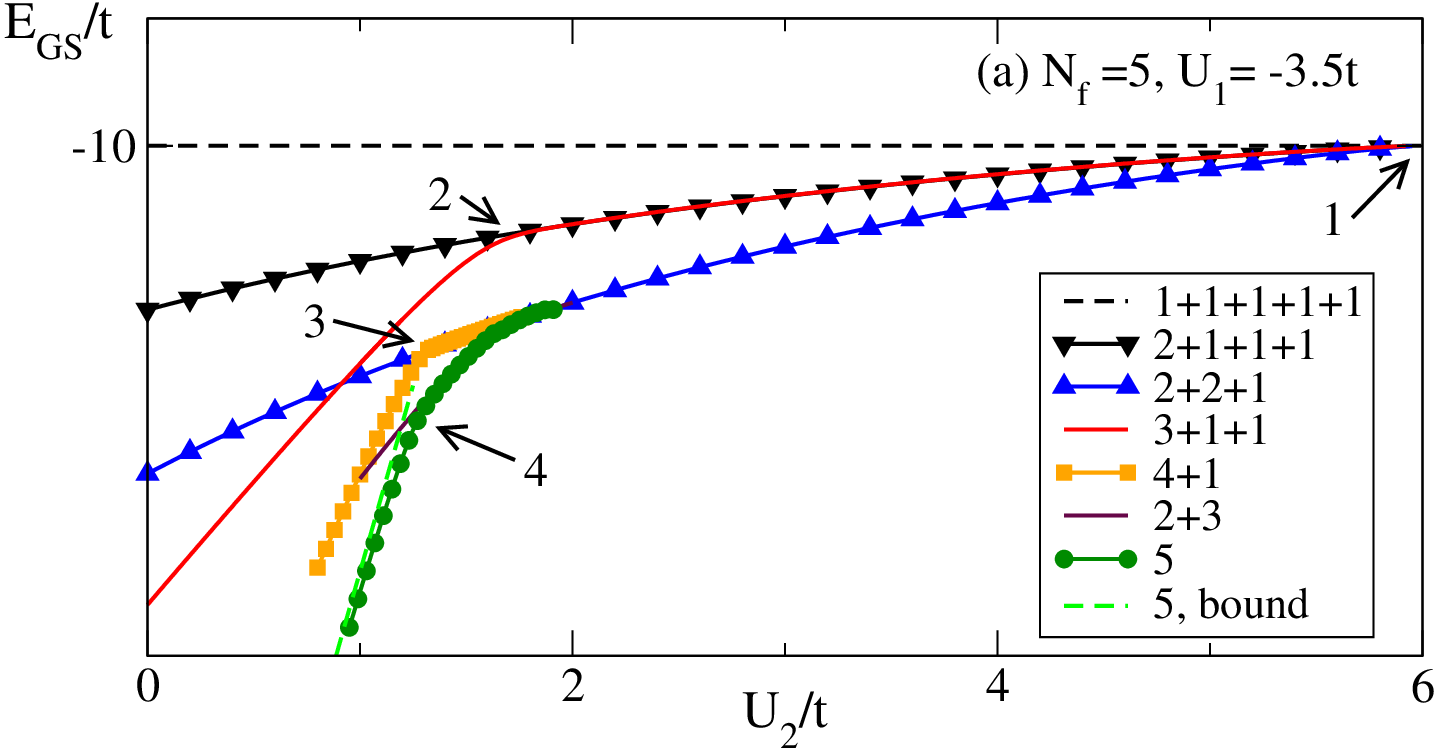}
\includegraphics[width=0.8\columnwidth]{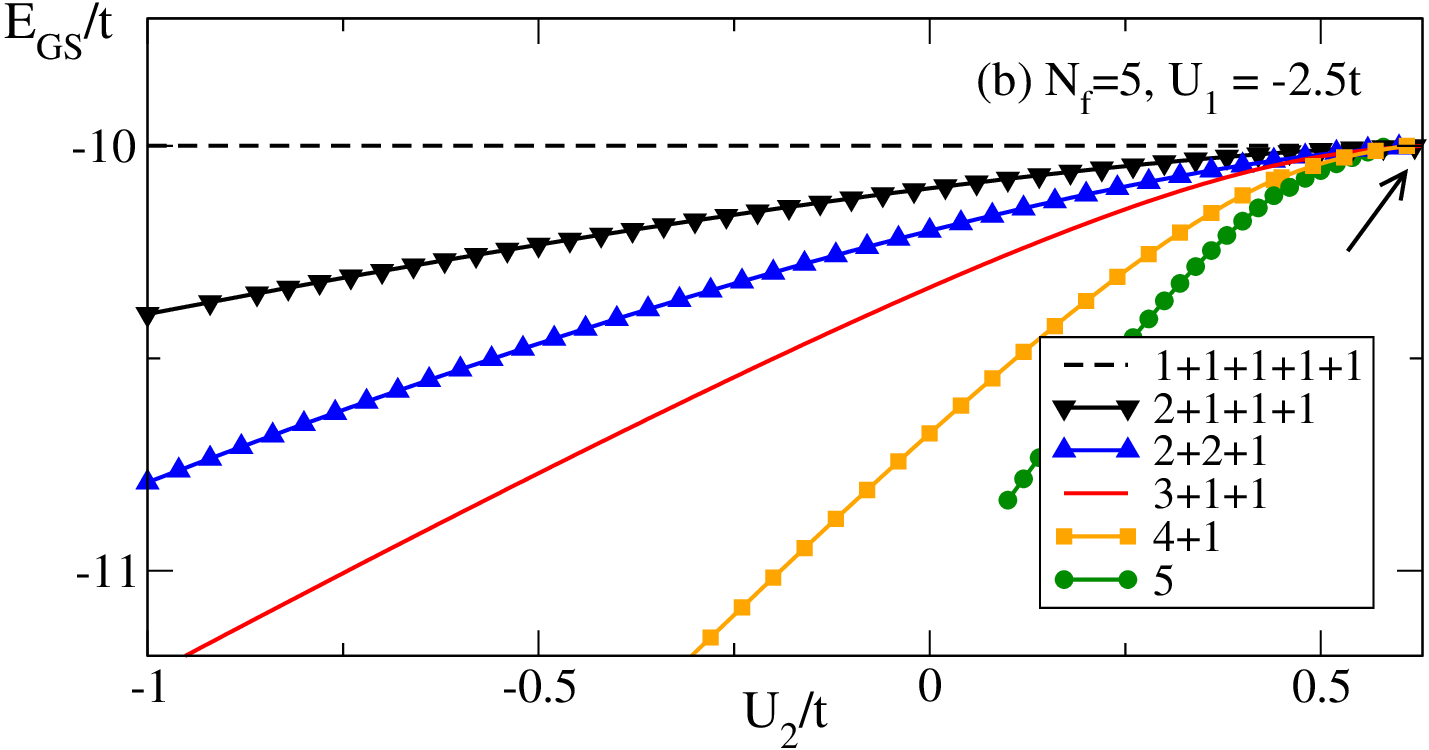}
\caption{(color online) $N_f=5$ GS energy (green circles, label ``5'')
  vs. $U_2$ for (a) $U_1=-3.5t$, and 
  (b) $U_1=-2.5t$. Other lines show the lowest energies of various
  complexes, and arrows indicate dissociations (see text for details). 
\label{fig3}}
\end{figure}

The fact that stable trions are found for $N_f=3$ does not, however,
guarantee that they appear at finite concentrations. Just as the
pair+fermion is unstable to trion formation, trions may be unstable to
bigger bound complexes, if more particles are present. Indeed, a study
of cases with $N_f=4$ and 5 fermions proves that trions are actually
unstable. This is shown in Fig. \ref{fig3}(a)  where
we plot the energy of the $N_f=5$ GS vs. $U_2$ (line marked
``5'') at $U_1=-3.5t$. The other lines show energies where a continuum could
appear, {\em eg.} $E_{2+2+1} = 2E_{2,GS}-2t$ is the lowest energy
of two pairs plus a fermion, $E_{2+3} = E_{2,GS}+E_{3,GS}$ is the
lowest energy for a pair plus a trion, etc. The arrows indicate various
dissociations. Arrow 1 shows when a pair becomes more stable than
2 fermions ($E_{2+1+1+1}< E_{1+1+1+1+1}$), while arrow 2 shows when a
trion becomes more stable than a 
pair+fermion ($E_{3+1+1} < E_{2+1+1+1}$), see
Figs. \ref{fig2}a,b. A trion+fermion is unstable to either
 two pairs (at larger $U_2$) or a 4-fermion bound complex
(smaller $U_2$). The boundary between the two is marked by
arrow 3 ($E_{4+1}=E_{2+2+1}$). But 4-fermion bound states are not
stable either, since  $E_{4+1} < min( E_{2+3}, E_{5})$ 
(arrow 4 marks where the 5-fermion bound complex 
breaks into a pair+trion). Below it, $E_5$ is indeed in good
agreement with the perturbational estimate for the energy of a 5-bound
complex $E_{5,B}=4U_1+3U_2+2t^2/(U_1+t^2/U_1-2t^2/(U_1+U_2))$, shown by the
  dashed line indexed ``5, bound''.

\begin{figure}[t]
\includegraphics[width=0.75\columnwidth]{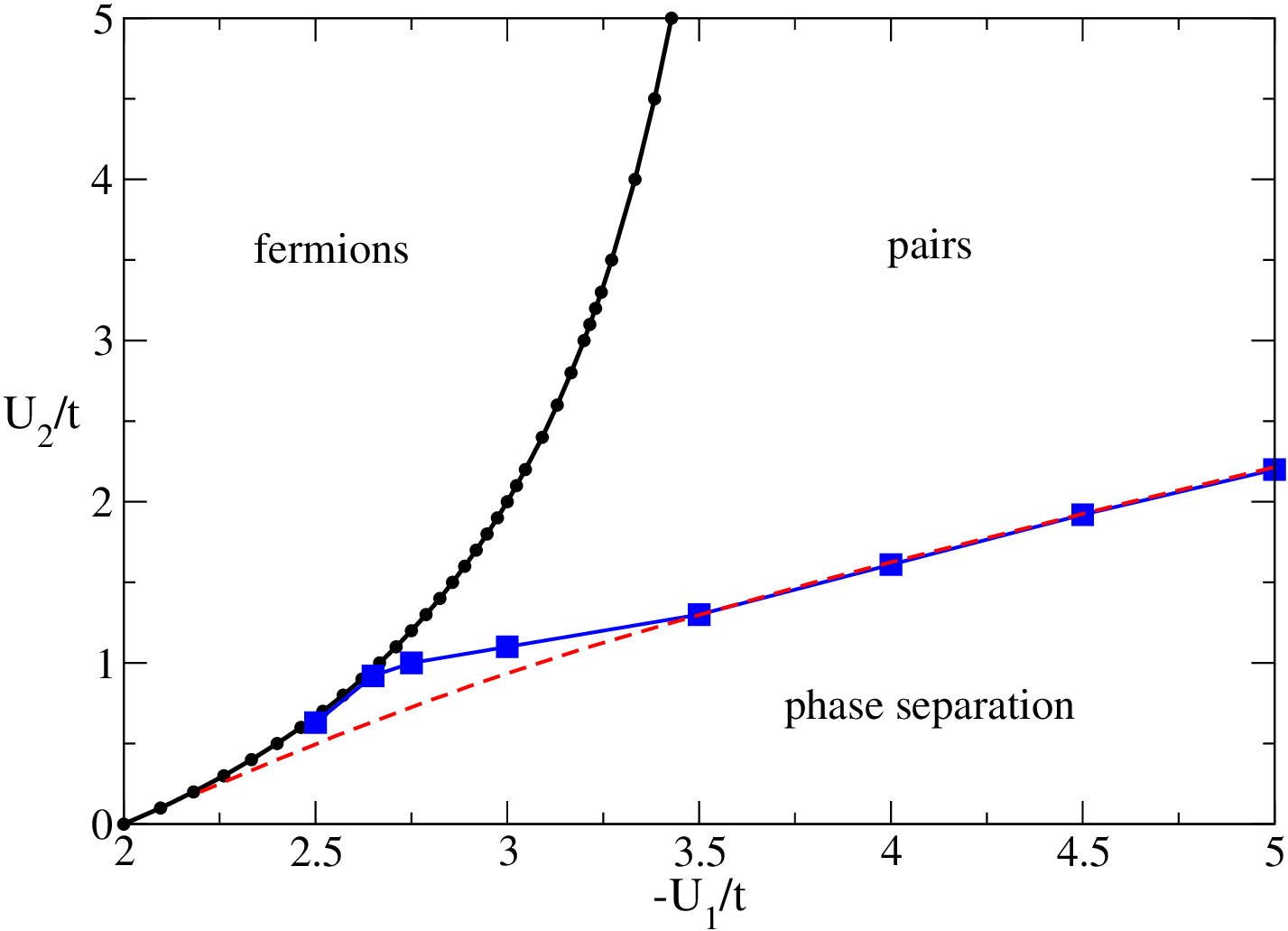}
\caption{(color online) Phase diagram at small concentrations. The GS consists
  either of unbound fermions, or of pairs, or
  it phase separates into fermion rich and fermion poor regions. The
  dashed line is an estimate for phase separation (see text).
\label{fig4}}
\end{figure}

What happens as $N_f$ increases becomes clear if we realize that
arrows 3 and 4 point to essentially the same $U_2$ value. If more
fermions are added, below this $U_2$ we expect a bigger and bigger
bound complex -- in other words, phase separation occurs and the
system splits into a fermion rich and a fermion poor region. Above
this, a gas of pairs is stable (plus one trion, if $N_f$ is odd). That
this inference is correct is verified by the following argument. This
critical value should be given by the condition that adding two more
particles to a fermion rich region (which changes energy by about
$2U_1+2U_2$, because of extra interactions) should be energetically
favorable to having a bound pair far away. From $2U_2 + 2U_1 <
E_{2,GS}$ we find $U_2=1.29t$ if $U_1=-3.5t$, in good agreement with
the value $U_2=1.3t$ pointed to by arrows 3 and 4.

Thus, based on these few-particles results, we can infer the phase
diagram of this model at small concentrations, shown in
Fig. \ref{fig4}. The dashed line shows the estimate discussed above,
accurate for large $U_1,U_2$ (at smaller $U_1$, $t$ comes into play
since the extra fermions need not be fully localized at the edge of
the fermion rich region). If $U_1 > -2.6t$, the transition is from
phase separation to unbound fermions as $U_2$ increases. This is
shown, for $U_1=-2.5t$, in Fig. \ref{fig3}b: here each bigger complex
is more stable than any smaller ones, if $U_2 < 0.63t$ (arrow). 

While we are not aware of numerical studies of this model, the good agreement 
with various asymptotic estimates as well as with known
results for spin-${1\over 2}$ Hamiltonians \cite{supp},
supports the accuracy of our results. This work
shows that even such a simple model has a 
rich behavior that can be uncovered with this method.

To summarize, we have shown how to calculate few-particle Green's
functions on an infinite 1D chain. The information obtained from them
sheds light on the stability of few-particle bound states. It also
illustrates the dangers of an insufficient analysis -- if we stopped
at $N_f=3$, we would conclude that trions are stable in a large region
of the parameter space, in this model. Analysis for larger $N_f$ shows
that addition of more particles leads to instability of trions, and
furthermore allows us to find the phase diagram for small
concentrations.

Although these results are for a 1D model, as discussed above this method  generalizes to higher-D if the hopping is
nearest-neighbor only. This opens the way to study the stability of trions and
bi-excitons in realistic lattice models. Such work is currently under way.

{\em Acknowledgments:} I thank I. Affleck, G. Sawatzky, P.  Stamp,
F. Zhou and S. Yarlagadda for useful discussions. This work was
supported by NSERC and CIfAR.


\section{Supporting material}

\subsection{Further support for this phase diagram}

As discussed in the main text, the phase diagram we derived from the
few-particle Green's functions is supported by the asymptotic
lines shown in Figs. 2a (inset) and Fig. 4.  We are not aware of any
numerical studies of this model that could be used for direct
comparison, although it bears mentioning that once $M_c$ is large
enough that convergence has been achieved (see discussion and examples
below), these results are exact --
there is no approximation involved in obtaining them.

However,  additional support for these results can be obtained
from studies of spin$-{1\over2 }$ Hamiltonians. Through a
Jordan-Wigner transformation \cite{JW}, the Hamiltonian for spinless
fermions studied here can be mapped into:
$$
{\cal H} =
\sum_{i}\left[-2t\left(S^x_iS^x_{i+1}+S^y_iS^y_{i+1}\right)+
  U_1S^z_iS^z_{i+1}+U_2S^z_iS^z_{i+2}\right].  
$$
The case  $ U_1<0, U_2>0$ corresponds to 
ferromagnetic nearest-neighbour Ising interaction and frustrating
anti-ferromagnetic next nearest-neighbour Ising interaction.

For $U_2=0$, the phase diagram of this model is well-known: it
consists of a ferromagnet for $|U_1|> 2t$ and a Luttinger liquid
otherwise \cite{Giam}. The ferromagnet corresponds to
phase-separation in the fermion language, since the magnetization is
linked to the density of fermions. The unpaired fermions phase is
reasonably linked to the Luttinger liquid, so this line of
the phase diagram agrees with other known results. 

While we could not find a study of the spin model with $U_2\ne 0$, it is
expected that an increase in $U_2$ frustrates the ferromagnetic phase
and eventually makes it unstable. We believe that the pairs phase is
a charge 2 Luttinger liquid, a sort of 1D version of a p-wave
superconductor.  Such phases are fairly well known  \cite{Giam,Sch}.  

\subsection{Details for the $N_f=2$ solution}

As discussed, because the Hamiltonian is invariant to translations it
is convenient to use the two-fermion states:
$$ |k, n\rangle = {1\over \sqrt{N}} \sum_{i}^{} e^{ik
  \left(R_i+{na\over 2}\right)} c_i^\dag c^\dag_{i+n}|0\rangle.
$$ The method can be trivially extended to systems with disorder
because using these translational states is not an essential
ingredient of the method.

For a finite-size chain with $N$ sites, in order to not double count
the states, we must restrict $1\le n \le
N_{max}$, where $N_{max}={N\over 2}$ if $N$ is even, respectively
$N_{max} = {N+1\over 2}$ if $N$ is odd. In other words, when
considering the two fermions on the closed ring, we take the distance
between them to be the shortest possible ``arc'', $n\equiv \min(n,
N-n)$.  Using $e^{ikNa}=1$,  it follows that for any $1\le n\le
N_{max}$
$$ |k, N-n \rangle = - e^{ik{Na\over 2}} |k, n\rangle.
$$

The equations of motion for the two-particle Green's functions were
derived in the main text. For  $m=1$, and using the
notation $G(1,n;k,\omega) \rightarrow a_n$, they are:
\begin{align} 
\label{4}
&(\omega +i \eta - U_1) a_1 + f(k) a_2 =1 \\ &(\omega + i \eta - U_2)
a_2 + f(k) (a_1 + a_3) = 0 \\ &(\omega + i \eta) a_n + f(k)
(a_{n-1}+a_{n+1}) = 0
\end{align}
for $3\le n < N_{max}$, and finally
\begin{equation}
\EqLabel{8} (\omega + i \eta) a_{N_{max}} + f(k) \left[1-
e^{-ika{Na\over 2}}\right]a_{N_{max}-1}=0
\end{equation}
A numerical solution is now trivial.

However, we can do better. First, note that for $ka=\pi$ we have
$f(k)=0$, and so the solution is:
$$
G(1,n;{\pi\over a},\omega)={\delta_{n,1}\over \omega+i\eta
-U_1}.
$$
 This is expected,
since $|{\pi\over a}, 1\rangle$ is an eigenstate of the full Hamiltonian with
energy $U_1$, as can be easily checked.

For any $ka\ne \pi$, the equations for $3 \le n \le N_{max}-1$
can be solved analytically. Their general solution is:
$$ a_n = \alpha z^n +  {\beta\over z^n}
$$ where $z$ and  $z^{-1}$ are the roots of the characteristic
equation: $
(\omega + i \eta) + f(k)\left( z+{1\over z}\right)=0$. We 
{\em choose} $z\equiv z(k,\omega)$ to be the root for which $|z|<1$. This is always
uniquely defined for any $\eta > 0$, since the product of the two
roots $$ z_\pm = {1\over 2} 
\left[-{\omega + i\eta\over f(k)} \pm \sqrt{\left({\omega + i
\eta\over f(k)}\right)^2-4}\right]
$$ is 1. The solution is now immediate. The last equation for $n=N_{max}$ fixes
the ratio of $\beta/\alpha$, the solution is then propagated down to
$n=2$, and one uses the first two equations to find $a_1$ and
$\alpha$, completing the solution.

These results have a very straightforward physical
interpretation. Note that the general recurrence equation for $3\le n
\le N_{max}$ is the same that describes {\em  two free fermions}. For a
given $k$, the two-fermion continuum spans the energies $\{
\epsilon_{k-q}+\epsilon_q\}_q=[-4t\cos{ka\over 2}, 4t\cos{ka\over 2}]
$, where $\epsilon_k = - 2t\cos(ka)$ is the free particle energy.

This explains why for energies outside this range $|\omega| \ge 2f(k)=
4t\cos{ka\over 2}$, we find the two roots $z, 1/z$ to be real (when
$\eta \rightarrow 0$), and such that  $|z|\rightarrow 
0$ as $|\omega|\rightarrow \infty$. This shows that here 
 the two-particle
Green's function decays exponentially with $n$, as
expected since  there
are no free two-particles eigenstates at these energies. The
exponentially increasing $ {\beta\over z^n}\sim
z^{N-n}$ part is a finite size effect: on a finite chain, the
inter-particle distance eventually decreases  as $n \rightarrow
N$. In the limit $N\rightarrow \infty$, this contribution must vanish,
in other words in the thermodynamic limit we must have $a_n = z
a_{n-1}=\alpha z^n$ as a purely exponentially decreasing function with
distance. This is consistent with the fact that the solution must be
insensitive to how we end the recurrence relation (what is the
boundary condition) as
$N\rightarrow \infty$.

For certain values of $U_1, U_2$, new
eigenstates may appear outside the free two-particle continuum. These
describe bound pairs, therefore we expect
their wavefunctions (and the associated Green's functions)
to decay exponentially with the distance $n$ between the
particles. This is fully consistent with the previous
discussion. In particular, since $|z|\rightarrow 0$ as $\omega
\rightarrow -\infty$, it follows that
lower-energy pairs are 
more tightly bound.

On the other hand, inside the free two-particle continuum, $|\omega| <
2f(k)$, we find $|z|\rightarrow 1$ as $\eta \rightarrow 0$. As a
result, the two contributions to $a_n$ become oscillatory functions,
underlying the fact that there are freely propagating two-particle
eigenstates at these energies. The interaction will be responsible for
scattering leading to phase-shifts, however at these energies the two
particles can propagate arbitrarily far from each other, if $\eta=0$.

Using a finite $\eta$ (which we are forced to do, for numerical
reasons), results in $1-|z|\sim \eta$, in other words there is slow
exponential decay at these energies as well, but now controlled by
$\eta$. Physically, this is because the finite $\eta$ introduces a
``life-time'' for these particles. As a
result, if the distances $n, N-n$ are very large compared to the
typical relative distance the two free particles can explore in their
lifetime $\tau \sim {1\over \eta}$, the probability for the pair to be
at such distances decreases exponentially, and so do the Green's
functions.  It follows that in the thermodynamic limit, we can again
take $a_n = \alpha z^n$.  Then, the solution in the
thermodynamic limit is trivial since we only need to solve the
equations for $a_1$ and $a_2$ using $a_3=z(k,\omega)a_2$. The
solution is listed in the main text. Generalization to longer (but
finite) range interactions is trivial, as is finding the solution for
other $m$ values.

\subsection{$N_f=3$ in the thermodynamic limit}

Based on the arguments discussed above, in the
limit of an infinite chain we expect the Green's functions to decay
exponentially at all 
energies, with an exponent controlled by $\eta$ inside the
three-particle continuum, and by the inverse of the distance between
$\omega$ and the continuum's band-edge, for energies outside the continuum.

We use three-particle states of total momentum $k$:
$$ |k, n_1, n_2\rangle = {1\over \sqrt{N}} \sum_{i}^{} e^{ikR_i}
c^\dag_{i-n_1} c^\dag_i c^\dag_{i+n_2}|0\rangle
$$ where $n_1\ge 1, n_2\ge 1$. Since we take $N\rightarrow
\infty$, the states with $n_1\sim N_{max}, n_2\sim N_{max}$ become
irrelevant and we do need to worry about properly counting them;
because of the finite lifetime, even inside the 
continuum particles cannot travel that far from each other. For a
finite chain, however, proper counting  is 
important (see below).

The equations of motion for the three-particle Green's functions
defined in the main text, 
are:
\begin{widetext}
\begin{align*}
\EqLabel{12} \delta_{n_1,m_1} \delta_{n_2,m_2}= (\omega + i \eta
-U_{n_1, n_2}) G(m_1, m_2; n_1, n_2; k, \omega) \\ {+t \left[G(m_1,
m_2; n_1-1, n_2; k, \omega) + G(m_1, m_2; n_1+1, n_2; k, \omega)
\right.}\\ {+e^{ika} G(m_1, m_2; n_1-1, n_2+1; k, \omega) +
e^{-ika}G(m_1, m_2; n_1+1, n_2-1; k, \omega) } \\+ \left.G(m_1, m_2;
n_1, n_2+1; k, \omega) + G(m_1, m_2; n_1, n_2-1; k, \omega) \right].
\end{align*}
\end{widetext}
Here, $U_{n_1, n_2} = U_1 ( \delta_{n_1,1} + \delta_{n_2,1}) + U_2 (
\delta_{n_1,2} + \delta_{n_2,2}+\delta_{n_1+n_2,2})$ is the
interaction energy when the three particles are at relative distances
$n_1, n_2$ from each other. The remaining terms describe the effect of
hopping on the $|k, n_1, n_2\rangle$ state. The first two terms come
from the hopping of the left-most particle, which changes $n_1$. The
next two terms come from the 
hopping of the central particle, which keeps $n_1+n_2$ constant, 
and the last two terms are from the hopping of
the rightmost particle, which changes $n_2$. If $n_1=1$
then $G(m_1, m_2; n_1-1, 
n_2, k,\omega)\equiv 0$ since this hopping process is not allowed
for spinless fermions,
and similarly for $n_2$. For $N\rightarrow \infty$ we 
need not worry what 
happens as $n_1, n_2\sim N_{max}$, since the Green's functions vanish
before the particles go so far from each 
other.

Suppose we are interested in $m_1=m_2=1$ and use the shorthand
notation $a(n_1,n_2) = G(1,1;n_1,n_2;k,\omega)$. The resulting
infinite (in the thermodynamic limit) system of coupled recurrence
relations can be solved as follows. We define the vectors
$$ V_M = \left(
\begin{array}[c]{c}
a(1,M-1) \\ a(2,M-2) \\ \dots \\ a(M-1,1) \\
\end{array}
\right)
$$ which collect all the Green's functions with the same ``relative
distance'' $M= n_1+n_2$. Its dimension is $M-1$, although for
special values of $k$ there are further symmetries that can lower
it. For example, at $k=0$ we have $a(n, M-n) = a(M-n, n)$ and the
dimension is halved.

The special property of Hamiltonians with only nearest-neighbor
hopping is that the resulting equations of motion only link three
vectors with  consecutive 
relative distances. In other words, for any $M\ge 3$, we can
recast the recurrence equations as:
\begin{equation}
\EqLabel{rec} \gamma_M V_M = \alpha_M V_{M-1} + \beta_M V_{M+1}
\end{equation}
where $\alpha_M, \beta_M, \gamma_M$ are very sparse
matrices whose matrix elements are simple functions of $k,\omega$ that
can easily be read off the equations of
motion. Because we know that all Green's functions must vanish in the
limit $M\rightarrow \infty$, the solution of this recurrence equation
is given by:
$$ V_M = A_M V_{M-1}
$$ where the matrices $A_M$ are given by continued fractions:
$$ A_M = [ \gamma_M - \beta_M A_{M+1}]^{-1} \alpha_M
$$ and can be calculated starting with $A_{M_c+1}=0$ at a sufficiently
large cutoff $M_c$. 

Once all these matrices are known, and in particular $A_3$ which links
$a(1,2)$ and $a(2,1)$ to $a(1,1)$, we can use the equation of motion
with $n_1=n_2=1$ to find:
$$ G(1,1;1,1;k,\omega) = \frac{1}{\omega + i \eta - 2U_1 - U_2 +
t\left[A_3|_{1,1} + A_3|_{2,1}\right]},
$$ from which we can then get all the other propagators.

To illustrate the effect of the numerical parameters $\eta$ and $M_c$, we
analyze the three-particle spectral weight:
\begin{equation}
\EqLabel{t1}
A_3(\omega) = -{1\over \pi} \mbox{Im} G(1,1; 1,1;k=0,\omega)
\end{equation}
This is finite for all energies $\omega$  in the $k=0$, $N_f=3$
spectrum, and its weight gives the probability of
having the three fermions located on three consecutive sites. 

Just like for $N_f=2$, whether a bound trion is
the ground-state or not is determined by whether a discrete Lorentzian
appears below the continuum, or not. The continuum starts at
$E_{2,GS}-2t$, where $E_{2,GS}$ is the GS energy of the $N_f=2$ case. If
the parameters are such that bound pairs are not stable, then
$E_{2,GS}=-4t$ and the three-particle continuum starts at
$-6t$. However, if a bound pair is stable, then the continuum moves to
lower energies, and consists of states where a free particle scatters
off a bound pair (other higher-energy features are also
present, but not of interest for our analysis).

In Fig.~(\ref{fig3}), we show $A_3(\omega)$ for  $U_1=-3t, U_2=0$ and
three sets of parameters $\eta, N_c$. The dashed vertical line shows
the expected on-set of the 
continuum, at $E_{2,GS}-2t$ [for these parameters,  $E_{2,GS} \approx
-4.33t$, see Fig. 1c in the main text]. Clearly, $A_3(\omega)$ shows
a continuum starting at this energy, but there is also a Lorentzian peak
below it, indicating a stable trion for these parameters.

Note that the spectral weight in the continuum depends on the
specific broadening $\eta$ and cutoff $M_c$ used. This dependence can
be understood easily. $M_c$ is the cutoff at which we set the Green's
functions to zero. Physically, this is equivalent with adding an effective
``interaction'' which becomes infinite if the total relative distance between
particles $n+m > M_c$, and is zero otherwise. As is the case for any
system in a ``box'', we expect the continuum to be
replaced by a set of discrete levels, with a spacing $\delta E \sim
1/M_c$. These states, however,  are broadened by $\eta$. This explains
why the first curve is much smoother than the second one, even though
they have the same $M_c$. On the other hand, the third curve has $M_c$
increased by a factor of two, and indeed there are roughy twice as
many  oscillations marking the
discrete peaks. For any value of $M_c$, the curve becomes smooth if
$\eta$ is large enough so that $\delta E \sim \eta$. As already
discussed, physically this means that the lifetime $\tau \sim 1/\eta$
is so short  
that the particles cannot travel up to the boundaries of this potential
``box'' defined by $M_c$.

\begin{figure}[t]
\includegraphics[width=\columnwidth]{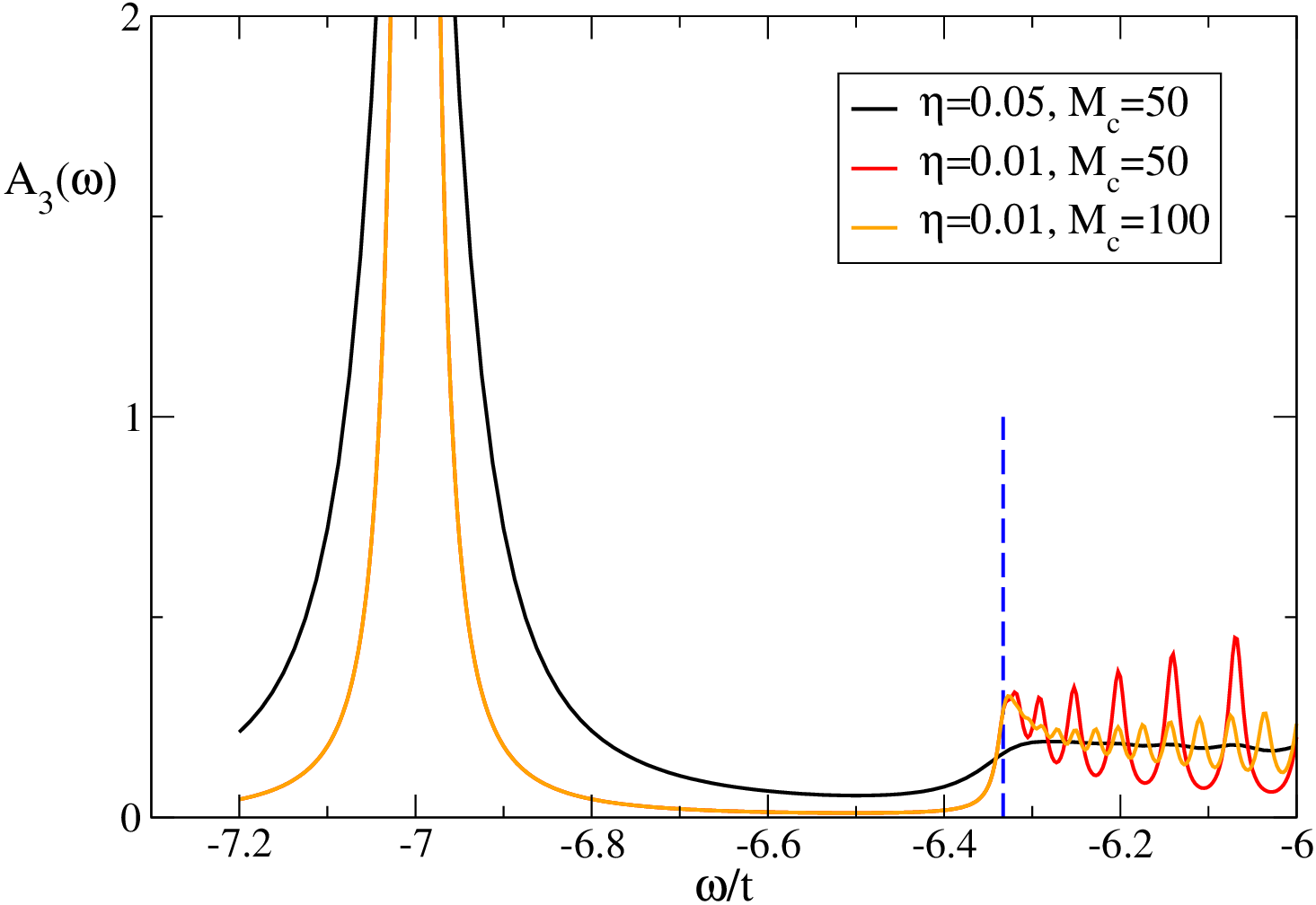}
\caption{$A_3(\omega)$ vs $\omega$ for $U_1=-3t, U_2=0$, for various
  values of the broadening $\eta$ and the cutoff $M_c$. The dashed
  line shows the expected continuum onset at $E_{2,GS}-2t$.
\label{fig3}}
\end{figure}

Below the continuum, the spectral weight is insensitive to $M_c$,
because the states that appear here (if any) are bound well inside
this ``box''. The broadening $\eta$ is still reflected in the shape of
the Lorentzian: although not shown entirely in Fig. \ref{fig3}, the
peak for the smaller $\eta$ is 5 times narrower and taller, as
expected. This insensitivity to $M_c$ is very convenient, because it
means that one can get very good estimates for the energy of strongly bound
states using rather small $M_c$ values. Of course, if the binding
energy is very small, then one has to increase $M_c$ until convergence
is achieved.

\subsection{Higher  $N_f$ in the thermodynamic limit}

It should now be apparent that the method generalizes
straightforwardly to any case with an odd number $N_f$ of
particles. We choose the reference location $i$ as being that of the
central particle, and index states in terms of the absolute values of
the relative distances of all other particles $n_1, n_2, ...,
n_{N_f-1}$ with respect to the central particle. In the equation of
motion, hopping of any of these other particles will increase or
decrease its own $n_i$ by 1, so the ``relative distance'' $M =
\sum_{i=1}^{N_f-1} n_i$ varies by 1. If the central particle hops to
the right, for example, this increases by 1 all the distances to all
particles to its left, and decrease by 1 all distances to all
particles to its right. Since there are equal numbers of particles to
the left and to the right of the central particle, then $M$ is
unchanged. 

For an even $N_f$, we choose as the reference particle either of the two
central particles. In this case, the ``relative distance'' is changed
by 1 when any of the particles hop, including the ``central''
one. 

\begin{figure}[t]
\includegraphics[width=\columnwidth]{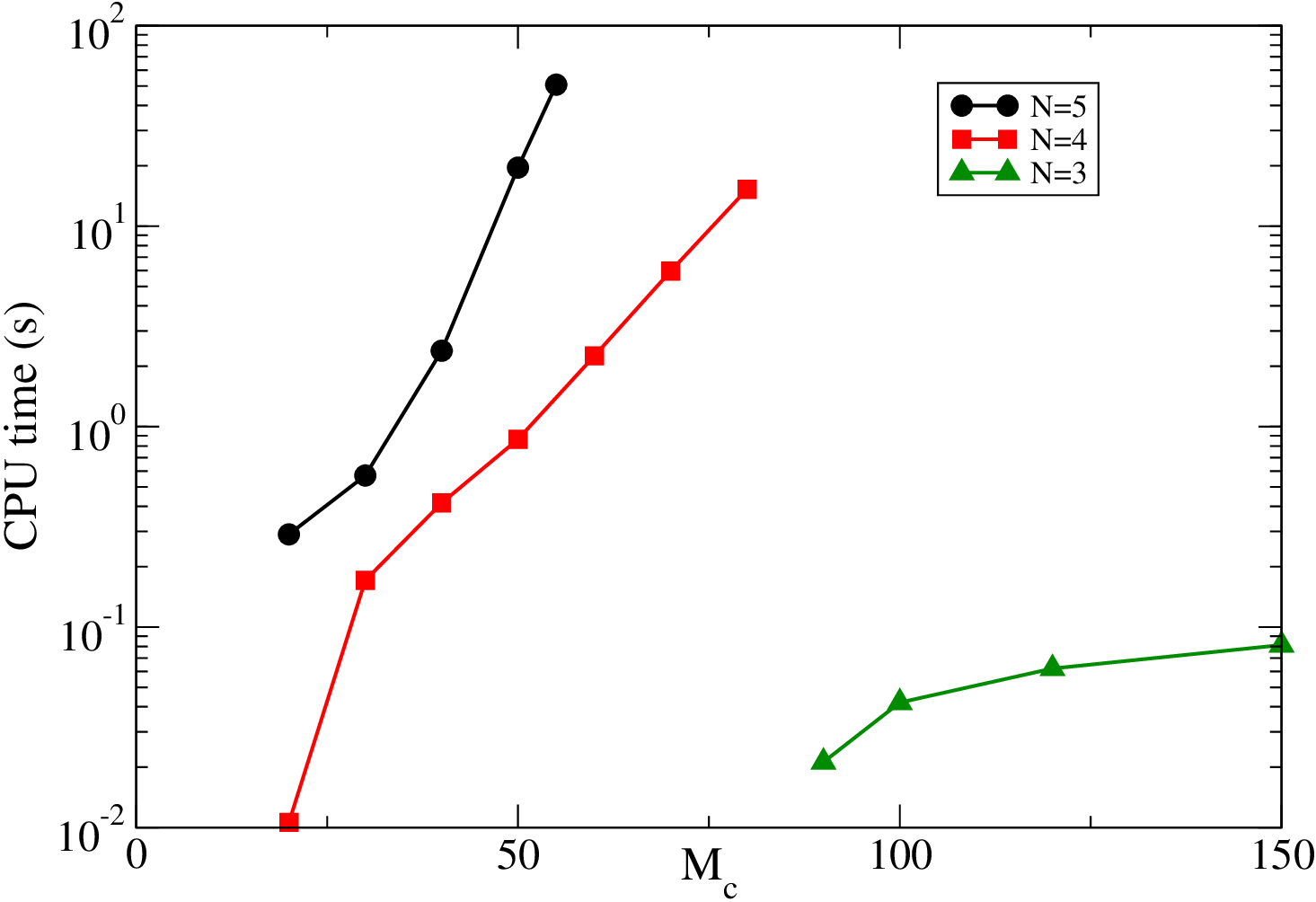}
\caption{CPU time for one frequency, vs. cutoff $M_c$, for systems
  with $N_f=3,4,5$.
\label{fig4}}
\end{figure}

In either case, the recurrence equation can 
still be cast in the general form of Eq. (\ref{rec}) and can be solved
by similar means. Of course, the larger $N_f$ is, the larger is the
dimension of $V_M$, so eventually one runs out of computational power
to calculate the continued fractions numerically. This is the factor
that limits what values of $N_f$ can be considered.

Fig. \ref{fig4} gives the real time to calculate the spectral weight
at one frequency on a 4-core CPU, for systems with $N_f=3,4$ and 5
fermions. As expected, the run times increase quite fast with both
$N_f$ and the chosen cutoff $M_c$. Note that for $N_f=3,4$ we showed
data for $M_c$ much larger than what is needed to achieve convergence,
simply because for smaller values the CPU time becomes independent of
$M_c$, showing that it is determined by other tasks, not by the
computation of the continued fractions which is the most
time-consuming part at large $M_c$. In fact, $M_c=50$ sufficed to
achieve convergence even in the most delicate cases discussed in the
main article, namely where a bound state is very close to a continuum
({\em i.e.} near a dissociation process). As already mentioned, here
one needs to use a small $\eta$ and therefore a larger $M_c$ to be
able to separate such close-by features. If the energy of the bound
complex is well below the continuum, on the other hand, much smaller $M_c$
suffices and calculations are much faster.

In practice, it is useful to first use a fairly small $M_c$ to quickly
scan a large range of energies to see where the main features are. Of
course, one particularly useful characteristic of this calculation is that the
spectrum for a given number $N_f$ of fermions must have one or more
continua at energies determined by the spectra with fewer fermions,
which are known. This gives not only a chance to validate the
computation, but also a very useful indication of where features are
expected in the spectral weight. If a $N_f$-bound complex is stable,
its corresponding Lorentz peak is below the lowest-energy continuum
and must be found by searching for a peak in the spectral weight in
the infinite range of energies lying below this lowest
continuum. Even in this case, one may use perturbation theory 
as a first guide to where the peak may be. As an example, see line
``5, bound'' in Fig. 3a which provides an estimate for the energy of
the 5-fermion bound complex. To zero order, its energy is
$E_{5,B}=4U_1+3U_2$, since a configuration with 5 fermions occupying
consecutive sites  has 4 nn and 3 nnn pairs. If $U_1, U_2$
are comparable to $t$, then one can use perturbation to
allow the end fermions to hop one site on and off the end of the complex;
this further lowers the energy by $2t^2/U_1$ (note that $U_1 < 0$),
since the resulting configuration only has 3 nn and 3 nnn pairs. If
needed, 2nd and higher order
corrections can be obtained by including further possible configurations,
in a standard fashion. Using such guidance plus low-$M_c$ scan of a
large range of energies, the rough position of the peak can be found
efficiently, after which $M_c$ 
is increased until the energy of the peak is converged to the
desired precision. Note that since this peak is a
Lorentzian with a known broadening $\eta$, as few as 2 points close to
its maximum suffice to extract its maximum and its weight from fitting.

This is why even though for $N_f=5$ and $M_c=50$ it
takes $\sim 20s$ to calculate the spectral weight at one frequency,
one can actually identify 
the GS energy with high precision within very few minutes, for a given
value of the parameters. This is also why we are confident that this
type of calculation can be successfully extended to larger $N_f$,
especially if clusters with more than 4 CPUs are used to further speed
up the computation. We stopped at $N_f=5$ here simply because this value
was sufficient to deduce the phase diagram for this model.

\begin{widetext}

\subsection{Mix of two different kinds of fermions}

We now briefly discuss the generalization to a mix of two different
types of spinless fermions, still on a 1D infinite chain. Let $c_i$
and $d_i$ be their corresponding annihilation operators. For
spin-${1\over 2}$ fermions, one can take $a_{i, 
  \uparrow} \equiv c_i; a_{i,\downarrow} \equiv d_i$. The model
Hamiltonian we consider is a direct generalization that used in the
main text:
\begin{multline}
\EqLabel{m1}
{\cal H} = -t_c\sum_{i}^{} (c_i^\dag c_{i+i} + h.c.) + U_{1,c}
\sum_{i}^{} n_{c,i} n_{c,i+1} + U_{2,c} \sum_{i}^{} n_{c,i} n_{c,i+2}
\\
-t_d\sum_{i}^{} (d_i^\dag d_{i+i} + h.c.) + U_{1,d}
\sum_{i}^{} n_{d,i} n_{d,i+1} + U_{2,d} \sum_{i}^{} n_{d,i}
n_{d,i+2}\\
+ U_0 \sum_{i}^{} n_{c,i} n_{d,i} + U_{1,m}
\sum_{i}^{}\left( n_{d,i} n_{c,i+1} +h.c. \right)+ U_{2,m} \sum_{i}^{}
\left(n_{d,i} n_{c,i+2} + h.c.\right).
 \end{multline}
 \end{widetext}
In other words, each species has both nn and nnn interactions, while
the mixed interactions are on-site, nn and nnn. Of course, if the two
spinless fermions correspond to different spin-projections of the same
spinful fermion, and the interactions are spin-independent, then one
expects $U_{1,c}=U_{1,d}=U_{1,m}$ etc. 

Consider first a mixed pair with a total momentum $k$. One expects to
be able to factorize the two-particle states into analogs of
``singlet'' and ``triplet'' (with $m=0$) states, which should not mix
with one another through hopping, so that each should have its own set
of recurrence relations. This is indeed true, however at
finite momentum these symmetries get mixed and identifying the proper
states requires a bit of work. The solution is as follows. 
We define:
$$
t(k) = t_c e^{i{ka\over 2}} + t_d e^{-i{ka\over 2}}=T(k) e^{i \phi_k} 
$$
where $T(k) = \sqrt{t_c^2+t_d^2 +2 t_c t_d \cos(ka)}$ and $\cos
\phi_k = (t_c+t_d)\cos{ka\over 2} / T(k)$. Then, let:
$$
 |k,s,0\rangle = {1\over \sqrt{N}} \sum_{i}^{} e^{ikR_i} c_i^\dagger
  d_i^\dagger|0\rangle
$$
and for any $n\ge 1$,
\begin{widetext}
\begin{align}
|k,s,n\rangle = {1\over \sqrt{2N}} \sum_{i}^{} e^{ikR_i+{na\over 2}}\left(e^{in\phi_k} c_i^\dagger
  d_{i+n}^\dagger- e^{-in\phi_k} d_i^\dagger
  c_{i+n}^\dagger\right)|0\rangle\\
|k,t,n\rangle = {1\over \sqrt{2N}} \sum_{i}^{} e^{ikR_i+{na\over 2}}\left(e^{in\phi_k} c_i^\dagger
  d_{i+n}^\dagger+ e^{-in\phi_k} d_i^\dagger
  c_{i+n}^\dagger\right)|0\rangle
\end{align}
\end{widetext}
The ``s'' and ``t'' labels are because at $k=0$, and if
these are spin-up and spin-down fermions, these states describe the
usual singlet and triplet combinations. 

It is now easy to check that the recurrence relations do
not mix ``s'' and ``t'' states together, and each set can be solved
similarly to that for the spinless $N_f=2$ case. This is a double
bonus. First, because it keeps the recurrence equations simpler, which
makes the calculation more efficient. More importantly, one can figure
out the symmetry of the bound states that form, based on which
Green's functions exhibit poles at those energies.

\subsection{More mixed fermions}

Generalization to more particles follows closely. The main
ingredient, namely that a relative distance can be defined, and that
hopping only varies it by at most 1, stays the same. The complication
is that now particles of unlike type can pass by each other, so for
example for a three-particle calculation with relative distance
$M=n+m$, one generally has to include all states like $\sum_{i}^{} e^{ikR_i}
c_{i-n}^\dagger c_i^\dagger d_{i+m}^\dagger|0\rangle$, $\sum_{i}^{} e^{ikR_i}
c_{i-n}^\dagger d_i^\dagger c_{i+m}^\dagger|0\rangle$ and $\sum_{i}^{} e^{ikR_i}
d_{i-n}^\dagger c_i^\dagger c_{i+m}^\dagger|0\rangle$. Thus, the
dimension of the vectors with a given relative distance is bigger
than if the particles were identical. Again, careful
consideration of symmetries (especially at $k=0$) lowers the
dimension and make the calculation more efficient, besides providing
information on the symmetry of the bound states.

\subsection{Bosons}

The calculation can be carried over to bosons trivially. The main
difference is that one can place any number of bosons on the same
site, so there are additional recurrence equations describing states
with shorter relative distances than possible for fermions. The nn
hopping insures the same general structure of the recurrence
equations, and in fact for the terms where there are no multiple
bosons at the same site, the equations are identical with those for
fermions in similar configurations.

\subsection{Higher dimensions}

To obtain this, one has to combine together the idea of a ``relative
distance'', described above for multiple particles, with that of a
``Manhattan distance'' which we introduced in Ref. \onlinecite{MpA} to
show how to calculate single-particle Green's functions in higher
dimensions. For example, for two particles on a 2D square lattice, one
needs  two integers $\mb{n}=(n_x,n_y)$ to characterize the
relative distance between the two particles in states of the form
$|\mb{k},\mb{n}\rangle\sim \sum_{i_x,i_y} e^{i\mb{k}\cdot\mb{R}_{\mb
    i}} c^\dag_{\mb i} c^\dag_{\mb i+\mb n}|0\rangle$. Some
restrictions apply to the allowed values of $n_x,n_y$ so that double
counting is avoided. For example, we can choose $n_x \ge 0$; if
$n_x=0$ then only $n_y\ge 0$ is needed, while if $n_x >0$, $n_y$ can
take both positive and negative values \cite{MG}. 
 Nearest neighbor hopping will link the Green's
function for this ket to the ones corresponding to kets with $(n_x\pm
1, n_y)$ and $(n_x,n_y \pm1)$.  As a result,  here we should choose
$M=|n_x|+|n_y|$ as a sum of the relative 
distances projected along all the axes -- this is a ``Manhattan
distance'' characterizing the relative distance between the two
particles in this state.  Nearest-neighbor hopping then
preserves the general structure of linking Green's functions with a
given $M$ to only others with $M\pm 1$, and the general approach of
rewriting the equations of motions in terms of continued fractions
carries over. In fact, this problem is very similar in structure to
that of 3 fermions in 1D, where we also need two integers $n_1, n_2$ to
characterize each internal arrangement, and where $M=n_1+n_2$. The
main difference is that in 2D, $n_y$ could be negative as well, in
other words there are roughly twice as many states with a given $M$
then for the 3 particles in 1D. This suggests a corresponding increase
in the computational time. In reality, even this increase can be
eliminated, if one explicitly works with pairs of $s$-wave or $d$-wave
symmetry. In the former case, the Green's functions corresponding to a
given $n_x$ and $\pm n_y$ are equal, while in the latter case, they
have equal magnitude but opposite 
sign. In either case, the actual number of unknowns is halved. As  a
result, the 2D calculation for either an $s$-wave or $d$-wave  pair is
basically equivalent to a  1D
calculation for 3 particles.

The generalization to 3 particles is described briefly in the main
article. In this case, we need 4 integers to describe the relative
positioning with respect to the ``central'' particle (note that
different particles can play this role, for different axes). In 1D, 4
integers are needed to describe the relative arrangements of 5
particles. In general, $m$ particles in 2D require $2(m-1)$ integers
to specify relative positioning from a ``central'' particle, and as
such, this calculation maps onto a calculation with $2m-1$ particles
in 1D. Just as discussed above for $m=2$, at first sight there are
more states with the same total $M$ in the 2D problem then in the 1D
one, because some of the 2D integers could be negative while the 1D
ones are all positive.  However, if symmetries are properly enforced
this overall multiplication factor can be removed, at least at
$k=0$. As a result, one finds not only the spectrum but also the
symmetry of the bound complex (if, indeed, such a bound complex is
stable).

This is why in terms of running times for higher-D, a reasonable
estimate can be obtained based on running times in 1D. For example,
the 2D, $N_f=2$ case discussed above is roughly equivalent to a 1D,
$N_f=3$ computation, since in both cases the configurations are
characterized by 2 integers. It follows that stability for trions
($N_f=3$) and bi-excitons ($N_f=4$) can be studied very easily in 2D,
since it involves problems similar to $N_f=5$, respectively $N_f=7$ in
1D. In 3D, study of trions would be equivalent to investigating
$N_f=7$ in 1D, which can certainly be accomplished in a reasonable
time on a regular desktop with a 4-core CPU. To study bi-excitons in
3D (equivalent to $N_f=10$ in 1D), it may be needed to use a
bigger cluster to lower the computation time. Of course, if the bound
complex is stable, this will be confirmed by a small $M_c$ run, in an
efficient fashion.

\subsection{Longer-range hopping and/or finite chains}

In these cases, it is impossible to recast the equations of motion in
terms of recurrence equations for consecutive vectors $V_{M-1}, V_M,
V_{M+1}$. For longer range hopping, this is because the relative
distance will be changed by at least $\pm 2$ for second-nearest
neighbor hopping. Even for nearest-neighbor only hopping, in a finite
system this simple rule is broken for states with particles separated
by maximum allowed distances. The only exception is for $N=2$ on a
chain, where as discussed, the hopping from $n=N_{max}$ to $N_{max}+1$
is actually mapped into hopping to $N_{max}-1$, up to a phase
factor. In all other cases, the hopping out of these states with
maximally allowed relative distances will map into states which can
have quite different $M$ values.

One may still obtain a solution for such cases, by solving all of them
together as a linear system, instead of factorizing them into a
recursive set based on their $M$ values. While the dimension of this
linear system is much bigger, the matrix is extremely sparse and can
be dealt with efficiently by various known algorithms.

For an infinite system with longer range hopping, if one is
interested in bound states, then one can set a quite small cutoff
resulting in a reasonable computational task. For a finite-size chain,
one needs some physical intuition to decide what states (Green's
functions) can be removed from the calculation, i.e. how to define a
``cutoff''. We are currently investigating such problems.

\end{document}